\documentclass[twocolumn,amsmath,amssymb,aps,prl,superscriptaddress]{revtex4-1}
\usepackage{graphicx}
\usepackage{dcolumn}
\usepackage{bm}
\usepackage{float}
\usepackage{amssymb}
\usepackage{color}
\usepackage{multirow}
\usepackage{stmaryrd}
\usepackage{epstopdf}

\begin{document}

\title{Field-induced Metal-Insulator Transition in $\beta $-EuP$_3$}

\author{Guangqiang Wang}
\affiliation{International Center for Quantum Materials, School of Physics, Peking University, China}
\author{Guoqing Chang}
\affiliation{Laboratory of Topological Quantum Matter and Advanced Spectroscopy (B7), Department of Physics, Princeton University, Princeton, NJ, USA}
\author{Huibin Zhou}
\affiliation{International Center for Quantum Materials, School of Physics, Peking University, China}
\author{Wenlong Ma}
\affiliation{International Center for Quantum Materials, School of Physics, Peking University, China}
\author{Hsin Lin}
\affiliation{Institute of Physics, Academia Sinica, Taipei 11529, Taiwan}
\author{M. Zahid Hasan}
\affiliation{Laboratory of Topological Quantum Matter and Advanced Spectroscopy (B7), Department of Physics, Princeton University, Princeton, NJ, USA}
\affiliation{Lawrence Berkeley National Laboratory, Berkeley, CA, USA}
\author{Su-Yang Xu}
\affiliation{Department of Physics, Massachusetts Institute of Technology, Cambridge, Massachusetts 02139, USA}
\author{Shuang Jia}
\email{gwljiashuang@pku.edu.cn}
\affiliation{International Center for Quantum Materials, School of Physics, Peking University, China}
\affiliation{Collaborative Innovation Center of Quantum Matter, Beijing, 100871, China}
\affiliation{CAS Center for Excellence in Topological Quantum Computation, University of Chinese Academy of Sciences, Beijing 100190, China}
\affiliation{Beijing Academy of Quantum Information Sciences, Beijing 100193, China}
\begin{abstract}

Metal-insulator transition (MIT) is one of the most conspicuous phenomena in correlated electron systems \cite{RevModPhys.70.1039,Kuwahara961,CMRreview1}.
However such transition has rarely been induced by an external magnetic field as the field scale is normally too small compared with the charge gap.
In this paper we present the observation of a magnetic-field-driven MIT in a magnetic semiconductor $\beta $-EuP$_3$.
Concomitantly, we found a colossal magnetoresistance (CMR) in an extreme way: the resistance drops billionfold at 2 kelvins in a magnetic field less than 3 teslas.
We ascribe this striking MIT as a field-driven transition from an antiferromagnetic and paramagnetic insulator to a spin-polarized topological semimetal, in which the spin configuration of $\mathrm{Eu^{2+}}$ cations and spin-orbital coupling (SOC) play a crucial role.
As a phosphorene-bearing compound whose electrical properties can be controlled by the application of field,  $\beta $-EuP$_3$ may serve as a tantalizing material in the basic research and even future electronics.

\end{abstract}

\date{\today}
\maketitle

\begin{figure}[htbp]
\begin{center}
\includegraphics[clip, width=0.5\textwidth]{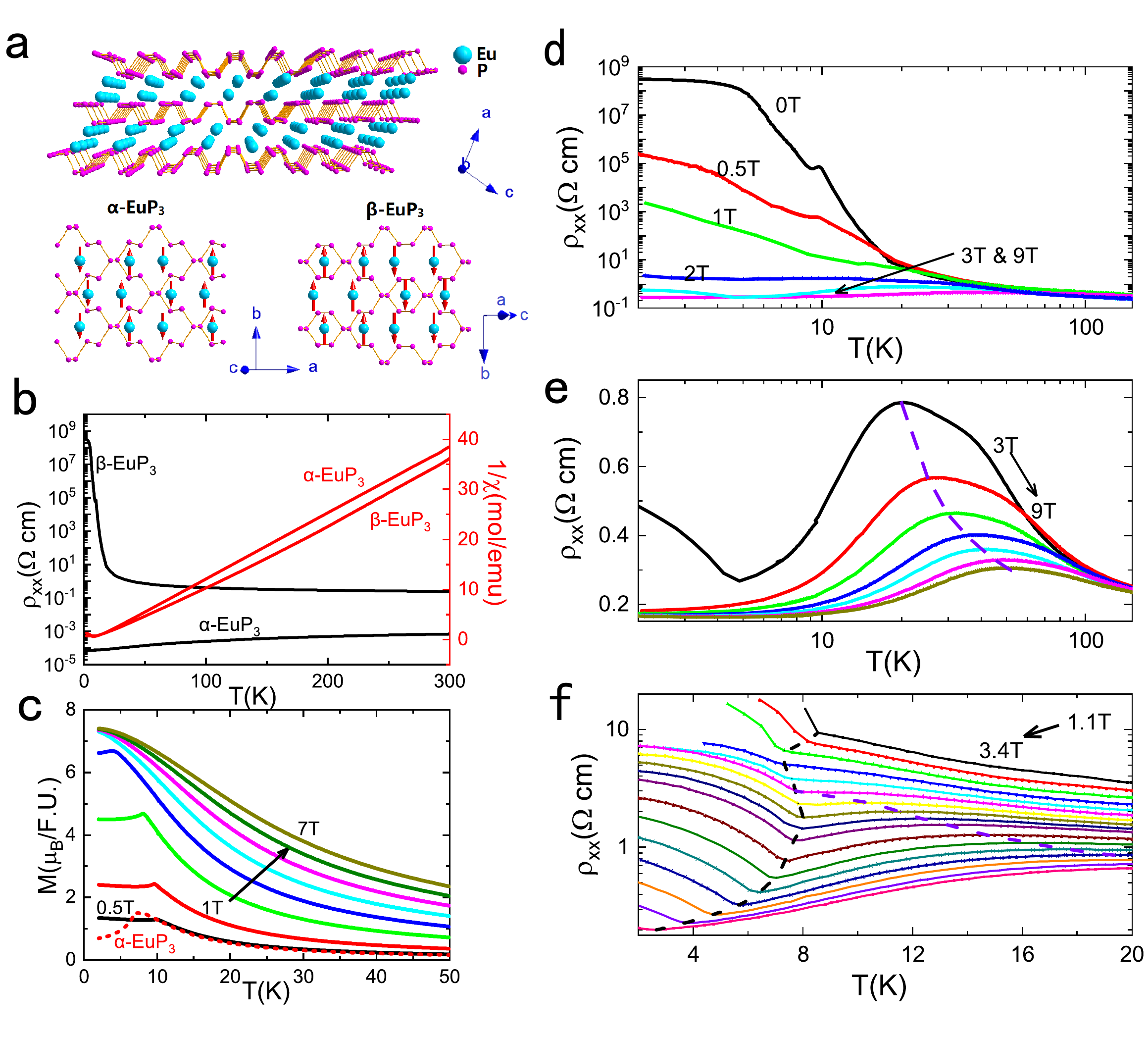}\\[5pt]  
\caption{Magnetic and electrical properties for $\alpha$ and $\beta $-EuP$_3$. \bf{a}, \rm The crystal and magnetic structure of AFM  $\alpha$ and $\beta $-EuP$_3$ in zero-field. The lower illustrations show the (001) projection of the phosphorus substructure and Eu$^{2+}$ cations layer. \bf{b}, \rm Temperature dependences of  inverse magnetic susceptibility [$1/(\chi - \chi_0)$] and zero-field electric resistivity ($\rho _{xx}$). \bf{c}, \rm Magnetization ($M$) under different external magnetic fields ($\mu _0H$) versus temperature. The solid lines are for $\beta $-phase while the dashed line is for $\alpha$-phase. \bf{d}, \rm $\rho _{xx}$ for $\beta $-phase under several representative $\mu _0H$ versus temperature. \bf{e}, \rm Zoom-in of the panel \bf{d} \rm demonstrates a broad peak on the $\rho_{xx} (T)$ curves in strong magnetic fields ($\mu _0H \geq  3~\mathrm{T}$). \bf{f}, \rm Zoom-in of the panel \bf{f} \rm in weak magnetic fields demonstrates complicated features of the $\rho_{xx} (T)$ curves including an reentry of the insulating state below the MIT. The $\mathcal{S}$-shaped black dashed line and the violet dashed line,  connecting the reentry points and the MIT points, respectively, are a guide to the eye. N. B. here the direction of the magnetic field is along the stacking direction (perpendicular to the phosphorus layer). For the data when the field is along the the phosphorus layer, see SI.}
\label{fig:1}
\end{center}
\end{figure}

Europium triphosphide (EuP$_3$) belongs to a family of AEPn$_3$ (AE = alkaline earth and Eu, Pn = P and As) compounds which are consist of divalent cations and  two-dimensional (2D) infinite puckered polyanionic phosphorus and arsenic layers \cite{bauhofer1982electrical, bauhofer1985electrical}.
The  AEPn$_3$ family crystallize in two closely related structures, named as $\alpha $ and $\beta$-phase, and both are thermaldynamicaly stable in EuP$_3$.
If we take a close look on a phosphorus layer in EuP$_3$, it can be generally derived from phosphorene by removing one quarter of the atoms so that it is built up of 14-membered phosphorus rings in $\alpha $-phase, but equal amounts of 6-membered and 22-membered phosphorus rings in $\beta$-phase [Fig. \ref{fig:1}(a)] \cite{bauhofer1985electrical, carvalho_phosphorene:_2016}.
This small change in the phosphorus layer leads to a striking difference in their electronic structures: $\alpha $-EuP$_3$ is a semimetal while $\beta$-phase is a semiconductor.
Figure \ref{fig:1}(b) shows different temperature dependent resistivity for $\alpha $ and $\beta$ phases: the former has a standard metallic profile, similar as observed in the isostructural SrAs$_3$ whereas the later shows continuous increases with decreasing temperature, leading to  $2\times 10^8~ \Omega~\mathrm{cm}$ at 2~K.

Topological electronic structure has been well-studied in non-magnetic $\alpha $-phase CaAs$_3$ and SrAs$_3$ \cite{quan2017single, xu2017topological, li2018evidence, Li_2019, SrAs3MR2019, SrAs3ARPES2020,hosen_experimental_2020, cheng_pressure-induced_2020}, but the electrical properties for the magnetic EuP$_3$ has not been fully investigated until now \cite{chattopadhyay1986incommensurate, schroder1986magnetic}.
Figure \ref{fig:1} (b) shows that the $\alpha $ and $\beta$ phases manifest very similar magnetic properties governed by the $^8S_{7/2}$ Hund's ground state of the $4f^7$ electron configuration of Eu$^{2+}$ cation.
The high-temperature susceptibility for both shows Curie-Weiss behavior ($\chi - \chi_0= \frac{N_A\mu _0\mu ^2_{\mathrm{eff}}}{3k_B(T-\theta _p)}$) with the effective moment $\mu _{\mathrm{eff}}=7.7~\mu _B/$formula unit (F.U.).
At low temperature both compounds undergoes a second-order phase transition from paramagnetic (PM) to antiferromagnetic (AFM) at Ne$\acute e$l temperature ($T_{\mathrm{N}}$) of about 10~K in zero-field [Fig. \ref{fig:1}(c)].
The magnetic structures for the AFM ground state are very similar as well: the $S_{7/2}$ local moments align along in the direction of the phosphorus plane to form FM planes in the sequence of one-in-one-out for $\alpha $-phase but two-in-two-out for $\beta$-phase \cite{chattopadhyay1988neutron} [Fig. \ref{fig:1} (a)].
This AFM ground state manifest relatively weak anisotropy and application of a field ($\mu _0H$) higher than $3~\mathrm{T}$ can polarize the moment along the stacking direction, leading to a saturated magnetization close to $7~\mu _B/$F.U., which indicates no valence change of Eu$^{2+}$ cation under magnetic field [Fig. \ref{fig:1} (c)].

\begin{figure}[htbp]
\begin{center}
\includegraphics[clip, width=0.5\textwidth]{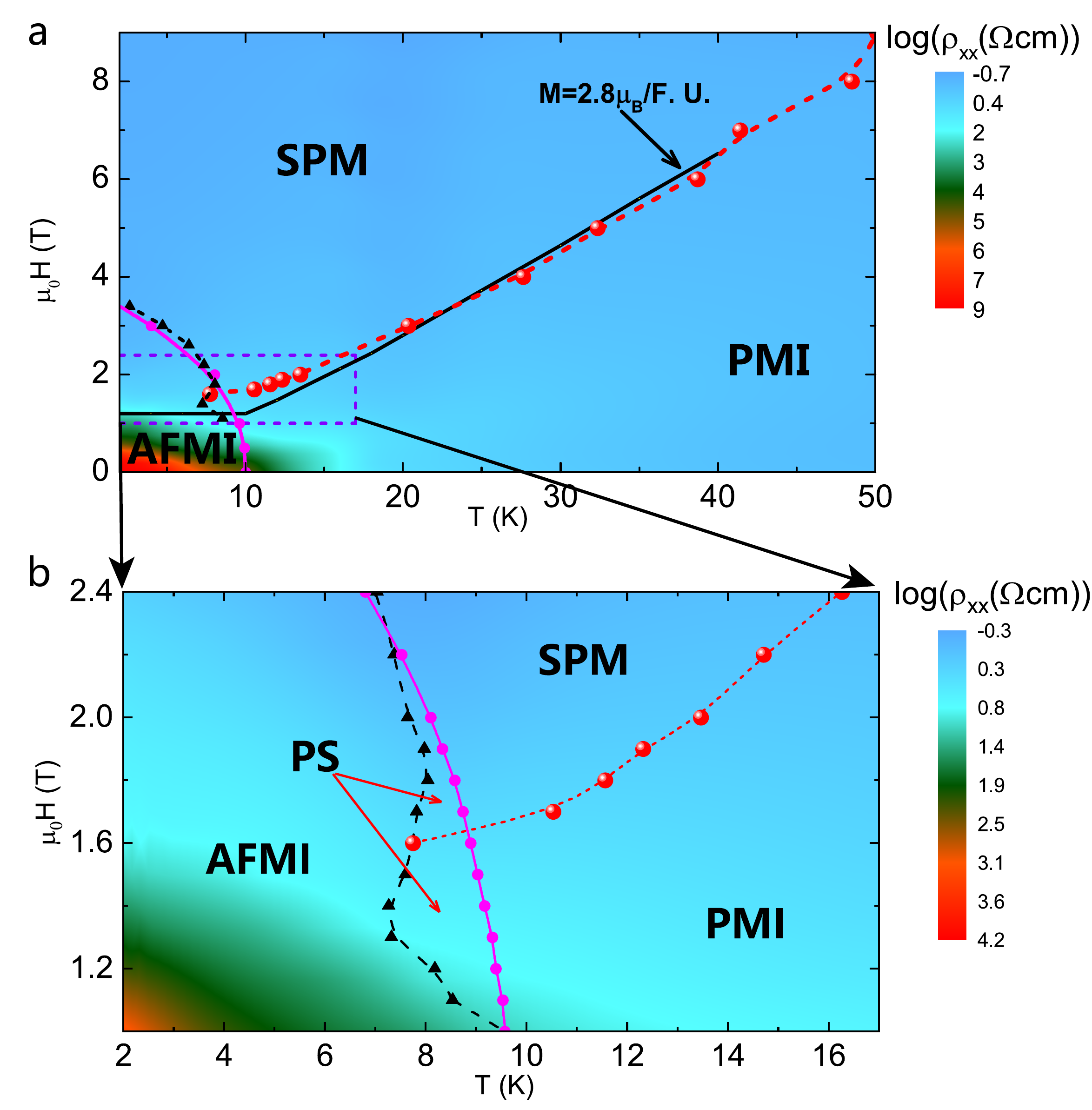}\\[5pt]  
\caption{Contour map in the $H-T$ plane for resistivity in $\beta $-EuP$_3$. The color bars at the side of the map represent the magnitude of the resistivity. The dots, squares and triangles denote the transition points. SPM, PMI and AFMI stand for the spin-polarized metal, paramagnetic insulator, and AFM insulator, respectively. \bf{a}, \rm a full diagram for $\mu _0H<9~\mathrm{T}$ and $T<50~\mathrm{K}$. The magenta solid line, determined by magnetization, is the boundary between AFM and SPM/PMI; the red dashed line is the boundary between PMI and SPM which approximately overlaps the iso-magnetization curve for $M=2.8~ \mu _B$/F. U. (black solid line). The black dashed line, determined by resistivity, is the boundary between AFMI and SPM. The black dashed line overlaps the magenta solid line below 7~K. \bf{b}, \rm Zoom-in of the phase diagram for  $1~\mathrm{T}<\mu _0H<2.4~\mathrm{T}$ and $2~\mathrm{K}<T<17~\mathrm{K}$ demonstrates a small saucer-shaped region of separation between magnetic and electric phases (labeled as PS).}
\label{fig:2}
\end{center}
\end{figure}

Our most conspicuous observation in $\beta$-EuP$_3$ is a field-driven metal-insulator transition (MIT), which can be seen in a series of $\rho _{xx}(T)$ curves under different magnetic fields in Fig. \ref{fig:1} (d).
As a consequence, the resistivity shows a colossal change in magnetic field at base temperature: it drops from more than $1\times 10^8~ \Omega~\mathrm{cm}$ in zero-field to less than $1~ \Omega~\mathrm{cm}$ in $3~\mathrm{T}$ at $2~\mathrm{K}$ (see the detail in SI).
This giga CMR in a moderate magnetic field easily dwarfs most of the large MR previously reported \cite{CMRreview1, tokura2006critical}.
As far as we are aware, only few perovskite-type manganese oxides show such huge reduction of resistivity in a magnetic field \cite{doi:10.1143/JPSJ.64.3626}.
For comparison, the semimetallic $\alpha$ phase shows 60\% of drop of MR in $9~\mathrm{T}$ at $2~\mathrm{K}$ (See SI).
It has been reported that the non-magnetic SrAs$_3$ shows a positive MR up to 80 in a large field at low temperatures \cite{li2018evidence,SrAs3MR2019,hosen_experimental_2020}.

Taking a close look on the $\rho _{xx}(T)$ curves we find a hump-like feature which is distinct from high-temperature semi-conductive and low-temperature metallic phases when $\mu _0H$ is higher than 3~T [Fig. \ref{fig:1} (e)].
We notice that the MIT boundary in PM state approximately overlaps the iso-magnetization curve of $M=2.8~\mu _B$/F. U. above 20~K in the phase diagram [Fig. \ref{fig:2}(a)].
Below 20~K, this part of MIT boundary slightly deviates the iso-magnetization curve and ends near the PM-AFM transition line about 10~K.
To nail down the MIT line below 10~K, we scrutinize the $\rho _{xx}(T)$ curves when $\mu _0H$ changes from $1.4~\mathrm{T}$ to $3.2~\mathrm{T}$ and find an apparent knee point in each $\rho _{xx}(T)$ curve which represents a reentry point to the insulating state at lower temperature [Fig. \ref{fig:1}(f)].
This reentry behavior has been observed in the MITs in many strongly correlated systems \cite{Limelette89}.

The phase diagram obtained from the magnetization and resistivity of $\beta$-EuP$_3$ contains rich features near the phase transition boundary lines [Fig. \ref{fig:2}(b)].
The reentry points in the $\rho _{xx}(T)$ curves overlap the PM-AFM transition below 8~K when  $\mu _0H$ is between $1.9~\mathrm{T}$ and $3.2~\mathrm{T}$ but slightly deviates from it at lower field, leading to a small saucer-shaped region of phase separation between the magnetic and electric phases.
Noticing the phase separation only occurs near the MIT boundary as long as $M$ is close to the critical value of $2.8~ \mu _B$/F. U., we attribute this feature to the thermal fluctuation which can be dominant in the vicinity of a phase transition in general.
The phase separation has been thought to play a critical role in the emergence of CMR in manganite and other materials \cite{dagotto_PS_2001}.

\begin{figure}[htbp]
\begin{center}
\includegraphics[clip, width=0.5\textwidth]{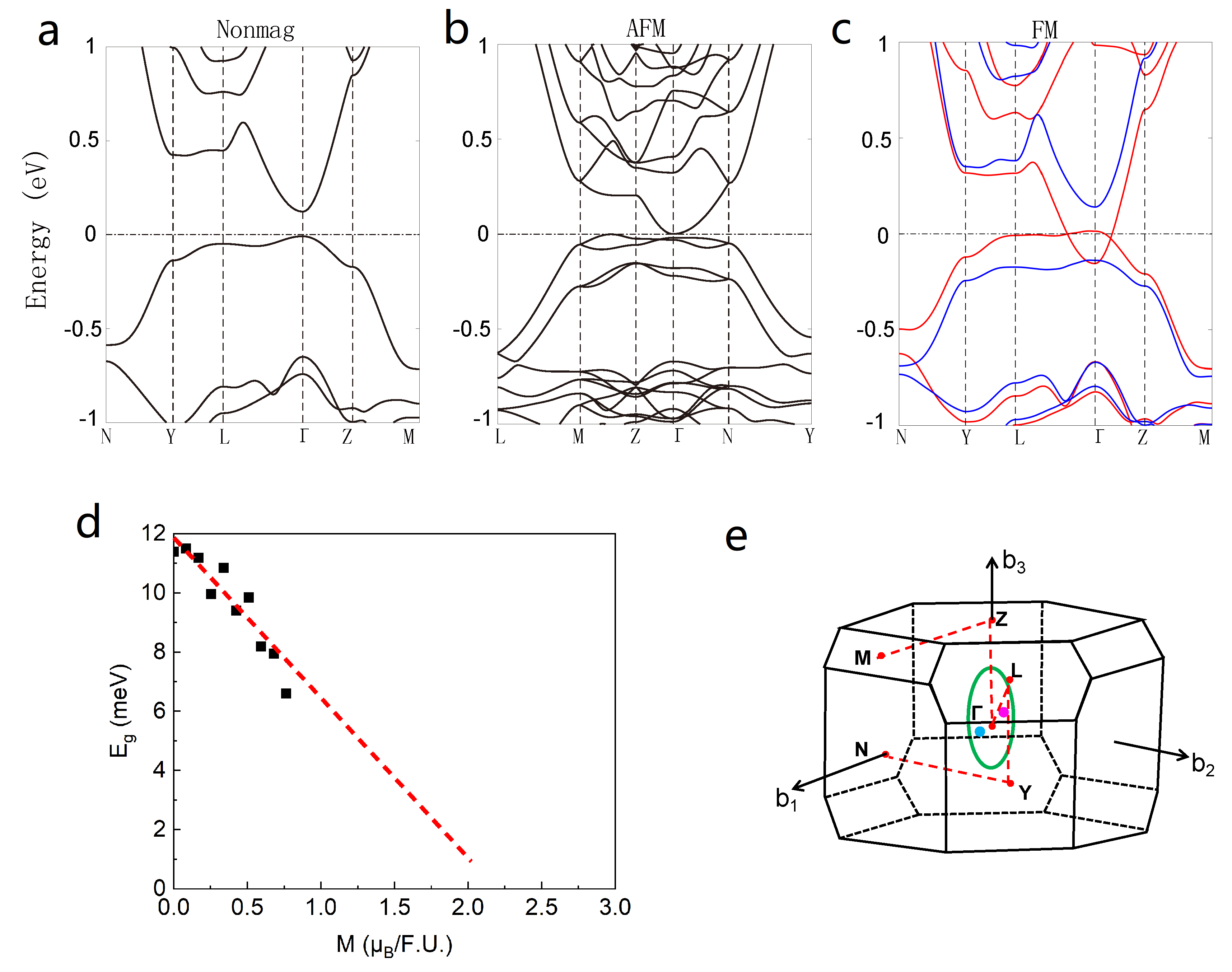}\\[5pt]  
\caption{ Electronic structure of $\beta $-EuP$_3$. \bf{a}, \rm fictitious nonmagnetic phase; \bf{b}, \rm AFM phase; \bf{c}, \rm fully SP (FM) phase. The red and blue lines represent spin-up and spin-down bands, respectively; \bf{d}, \rm band gap change with respect to the $M$. The result is obtained by fitting the $\rho _{xx}(T)$ curves from 100~K to 300~K with an activated form ($\rho =\rho _0e^{E_g/2k_BT} $). The dashed line is the linear extrapolation; \bf{e}, \rm locations of a nodal ring and a pair of Weyl points in the Brillouin zone. }
\label{fig:3}
\end{center}
\end{figure}

The phase diagram consist of paramagnetic insulator (PMI), spin-polarized metal (SPM) and antiferromagnetic insulator (AFMI), which are divided by the boundary lines determined by the hierarchy of interaction between magnetic ordering, spin polarization and thermal fluctuation.
Obviously there is no valence change in MIT while the spin texture must play a crucial role instead.
Our band structure calculation on a fictitious nonmagnetic $\beta$-EuP$_3$ shows a semiconductor which has a direct band gap at $\Gamma $ point [Fig. \ref{fig:3} (a)].
This semiconductive electronic structure is distinguished from the semimetallic $\alpha$-phase of nonmagnetic triphosphides and arsenides \cite{quan2017single, xu2017topological, li2018evidence}.
For the AFM ground state in zero-field, $\beta$-EuP$_3$ still has a small gap at $\Gamma $ point [Fig. \ref{fig:3}(b)].
By comparison, this band gap is fully closed in a spin polarized, FM state, leading to a band crossing forming two Weyl nodes in the $k_z=0$ plane and a nodal ring in the $k_x=0$ plane [Fig. \ref{fig:3} (c) \& (e)].
The minority spin band exhibits a $400~\mathrm{meV}$ gap while the spins of all electrons in the Weyl nodes and nodal ring are fully polarized, similar as what observed in the band structure in FM Weyl semimetal Co$_3$Sn$_2$S$_2$ and HgCr$_2$Se$_4$\cite{liu_giant_2018, wang_large_2018,Xu_HgCr2Se4}.

\begin{figure}[htbp]
\begin{center}
\includegraphics[clip, width=0.4\textwidth]{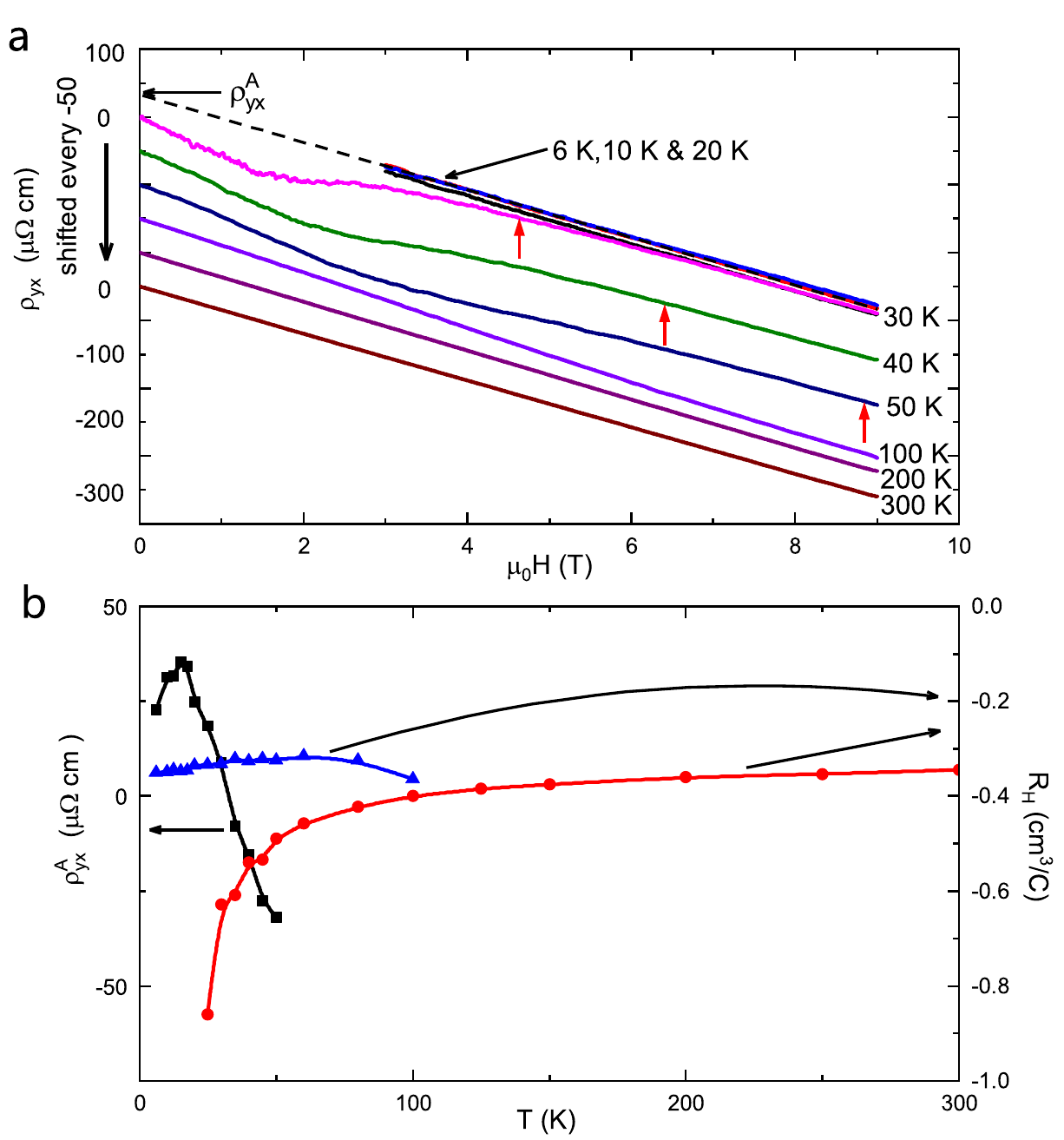}\\[5pt]  
\caption{ Hall effect of $\beta $-EuP$_3$.  \bf{a}, \rm $\rho _{yx}$ versus $H$ at representative temperatures. N. B. the $\rho _{yx}$ curves at $T = 40, 50, 100, 200$ and $300~$K are shifted $50~\mu \Omega~ cm$ more for each for clarity. $\rho _{yx}$ cannot be detected in low-field below 10~K as the resistivity is too high. The dashed line is the linear extrapolation of the high-field $\rho _{yx}$ which intercepts the anomalous Hall resistivity $\rho ^A_{yx}$ at $6$~K. The arrows indicate the MIT points at $30, 40$ and $50$~K. \bf{b}, \rm Parameters obtained from the fitting of $\rho _{yx}$. Black squares:  $\rho ^A_{yx}$. Red dots and blue triangles: the Hall coefficients $R_H$ in low- and high-field, respectively. The solid lines are a guide to the eye.}
\label{fig:4}
\end{center}
\end{figure}

Such spin-texture-induced band-closing scenario is evident by our analysis on the $\rho _{xx}(T)$ curves in different magnetic fields.
Fitting the $\rho _{xx}(T)$ curves from 100~K to 300~K with an activated form ($\rho =\rho _0e^{E_g/2k_BT} $) yields the band gap $E_g$ changing from 12~meV at zero-field to 6~meV at 9~T [Fig. \ref{fig:3} (d)].
Here $M$ is estimated as the value in the external field at 200~K.
Extrapolating the points we found that the gap will be close at $M=2.2~\mu _B$/F. U., very close to what we observe in the H-T phase diagram.
Moreover, the electronic structure change should be directly reflected in carrier density which will generate significant difference in Hall resistivity ($\rho _{yx}$) in low and high fields.
Figure \ref{fig:4} shows that the  $\rho _{yx}$ curves are linearly dependent on the field above $100~\mathrm{K}$.
Between $30~\mathrm{K}$ and $50~\mathrm{K}$, $\rho _{yx}$ apparently manifests a nonlinear behavior with two distinct slopes:  a larger slope in low field and a smaller slope in high field.
The arrows in Fig. \ref{fig:4} (a) demonstrate the transition points in the $H-T$ phase diagram and we find they match the slope change point in $\rho _{yx}$ exactly.
This nonlinear Hall effect is not due to the $M$ change because the $M(H)$ curves are nearly linear above 30 K, and we ascribe it to the transition from low-field semiconductor to high-field semimetal.
The Hall coefficient ($R_H$) of the low-field part shows a thermal excited feature [Fig. \ref{fig:4} (b)], which is consistent with the semiconductive $\rho _{xx}(T)$ curves and indicates that the thermal excited carriers take place in the electrical transport.
As comparison, the high-field part of  $\rho _{yx}$ is precisely parallel with each other at different temperatures and the high-field $R_H$ is nearly invariant up to 100~K, indicating that the band carriers, which is independent on the temperature, contribute to the electric transport in the SPM region.

We notice that the profile of $\rho _{yx}$ for $\beta $-EuP$_3$ is somehow reminiscent of that for manganites whose anomalous Hall effect (AHE) has been identified (See ref. \cite{Lyanda-Geller_AHE_2001} and the reference therein).
Extrapolating the high-field $\rho _{yx}$ at different temperatures, we find the intercept is not zero below 50~K.
If we write $\rho _{yx}=\rho ^A_{yx}+\rho ^N_{yx}=R_SM+R_HH$, the intercept represents an anomalous Hall resistivity $ \rho ^A_{yx}$ and the anomalous Hall coefficient $R_S$ equals  $ \rho ^A_{yx}/M$.
Figure \ref{fig:4} (b) shows that the $ \rho ^A_{yx}$ is positive below 20~K and the maximum value of $R_S$ equals $0.4~\mathrm{cm^3/C}$, close to the value of normal Hall coefficient $R_H$.
We compare this relatively large $R_S$ with the small ones in EuB$_6$ and EuS whose SOC is weak \cite{shapira1972resistivity, PhysRevLett.103.106602}.
Moreover, $\rho ^A_{yx}$ manifests a sign change at 30 K, which is far beyond the expectation based on the simple proportional relation $\rho ^A_{yx}\propto M$ and cannot be explained by a skew-scattering mechanism.
The sign change of $\rho ^A_{yx}$ with temperature has been observed in SrRuO$_3$ \cite{Fang92, PhysRevLett.93.016602} whose Berry curvature is very sensitive to the Fermi-level position and the spin-splitting.

The band-closing scenario can naturally explain the AHE in the SPM state of $\beta$-EuP$_3$.
Comparing the electronic structure of FM $\beta$-EuP$_3$ with the non-magnetic AEAs$_3$, we find that its Berry curvature is characterized by the Weyl nodes which are absent in the non-magnetic ones \cite{SrAs3ARPES2020,hosen_experimental_2020}.
This hot zone of the Berry curvature could be sensitively dependent on the gap closing and spin configurations.
The sign change of  $\rho ^A_{yx}$ may reflect the Berry curvature change near the MIT when $M$ is not fully saturated above 30 K.
If this band-closing scenario can be borne out by future study, the field-induced MIT in $\beta$-EuP$_3$ represents a relatively simple topological phase transition from a trivial semiconductor to a magnetic Weyl semimetal in which the Eu$^{2+}$ spin texture modifies the band alignment essentially.
It is known that the rare earth spin texture change can induce intriguing topological phase transition in rare earth half-Heusler compounds and iridate pyrochorides \cite{suzuki_large_2016, PhysRevLett.115.056402,tian_field-induced_2015}.
Here the resistivity and Hall effect change in $\beta$-EuP$_3$ is much dramatic.

In the above discussion, we have not considered the effect of vacancies which likely exist in the phosphorus layers in $\beta$-EuP$_3$.
It is well known that the vacancies in the europium-rich chalcogenides plays a crucial role on the CMR as they can generate extrinsic carriers and impurity band in these wide-band-gapped semiconductor \cite{oliver_EuO_1972,torrance1972bound,sinjukkow_EuO_2003}.
The exchange interaction between the $\mathrm{Eu^{2+}}$-bearing $4f$ moments and the conduction electrons induces a strong band splitting in $\mathrm{EuO}$ below its $\mathrm{T_C}$.
The electrons of the vacancies can propagate in the spin-polarized conduction band bottom when the vacancy states fall into the conduction band \cite{steeneken_EuO_2002}.
This picture seems to be able to explain the MIT in  $\beta$-EuP$_3$ as well, yet the existence of phosphorus vacancies and their role on the electronic structure need further elaboration.
Nevertheless, the field-induced MIT in $\beta$-EuP$_3$ highlights a transition from insulator to spin-polarized metal in a moderate magnetic field.
Our observation suggests an emerging and tantalising material which may have potential to bridge the research in the fields of spintronics and phosphorenics.

\section{Method}

Single-crystalline $\beta $-EuP$_3$ was grown directly by reacting stoichiometric europium pieces and red phosphorus powder in a vacuumed fused silica ampoule.
The ampoule was slowly heated to 900 $^oC$ and after 24 hours it was cooled to 750  $^oC$ at a rate of  $3^oC/h$.
Then the furnace was switched off and cooled to room temperature.
The ampule was cracked open with caution in ethanol in order to avoid ignition of residual white phosphorus when exposed to air.
Powder X-ray diffraction measurement confirmed the $\beta $-phase of EuP$_3$ (See Fig. S1 in SI).
The obtained crystals have an easy cleaving plane along (001) direction, as evidenced by X-ray Laue back-scattering experiment.

In order to make an ohmic contact between the semi-conducting sample and leads, Ti/Au contacts were deposited with  $10~\mathrm{nm}$ Ti layer and  $80~\mathrm{nm}$ Au layer by electron-beam evaporator, and then the sample was annealed in rapid thermal processing system at 500 $^\circ C$ for 100~s.
Platinum wires were loaded on the contacts with silver paste in a four-terminal method ready for magneto-transport measurement.
The contact resistance was determined as tens of ohms leading to linear I-V curves at room temperature.
The electric transport measurements were performed in a PPMS-9 cryostat (Quantum Design, Inc) with a base temperature of 1.8~K and magnetic field up to 9~T.
Because the capacity of the PPMS for measuring resistance in a four-point fashion is limited to the order of $\mathrm{M}\Omega $ magnitude, we employed Keithley 6221 as a DC current source of $10~\mathrm{nA}$ and 6430 as a voltmeter for measuring the resistance of the low-temperature insulating state.
For instance, the temperature-dependent resistance in zero-field above 8~K was collected through PPMS, while the data acquisition at lower temperatures was switched to 6221 and 6430.
The magnetization measurements were preformed in an MPMS-3 SQUID VSM (Quantum Design, Inc) from 2 K to 300 K.

\section{Acknowledgment}
S. Jia thanks for helpful discussion with J. H. Chen, T. L. Xia and Y. G. Shi.
We notice the work on $\mathrm{EuAS_3}$ from Cheng \it et. al. \rm \cite{cheng2020magnetisminduced} during the preparation of this draft.
The work at Peking University was supported by the National Natural Science Foundation of China No.U1832214, No.11774007, the National Key R\&D Program of China (2018YFA0305601) and the Strategic Priority Research Program of the Chinese Academy of Science (Grant No. XDB28000000).
Experimental and theoretical work at Princeton University was supported by the Gordon and Betty Moore Foundation (Grant No. GBMF4547 and GBMF9461/Hasan)

\bibliographystyle{unsrt}%

\clearpage
\end{document}